\documentclass[longauth]{aa}  

\usepackage{graphicx}

\usepackage{txfonts}

\usepackage{epstopdf}

\RequirePackage{color}

\usepackage[]{natbib}
\newcommand{\snrg}{\object{G5.9+3.1}}

\newcommand{\emaila}{donic@matf.bg.ac.rs}

\newcommand{\D}{$^\circ$}

\begin{document} 

\title{Murchison Widefield Array and \textit{XMM-Newton} observations of the Galactic supernova remnant \snrg}

\titlerunning{MWA and \textit{XMM-Newton} observations of the \snrg}

\authorrunning{Oni\' c et al.} 

\author{
D.~Oni\' c\inst{1}
\and
M.~D.~Filipovi\' c\inst{2}
\and
I.~Boji\v ci\' c\inst{2}
\and
N.~Hurley-Walker\inst{3}
\and
B.~Arbutina\inst{1}
\and
T.~G.~Pannuti\inst{4}
\and
C.~Maitra\inst{5}
\and
D.~Uro\v sevi\' c\inst{1,17}
\and
F.~Haberl\inst{5}
\and
N.~Maxted\inst{2,6} 
\and 
G.~F.~Wong\inst{2,6} 
\and 
G.~Rowell\inst{7}
\and 
M.~E.~Bell\inst{8}
\and 
J.~R.~Callingham\inst{9}
\and
K.~S.~Dwarakanath\inst{10}
\and 
B.-Q.~For\inst{11}
\and 
P.~J.~Hancock\inst{3}
\and 
L.~Hindson\inst{12}
\and 
M.~Johnston-Hollitt\inst{3}
\and
A.~D.~Kapi\' nska\inst{11}
\and 
E.~Lenc\inst{13}
\and 
B.~McKinley\inst{14}
\and 
J.~Morgan\inst{3}
\and 
A.~R.~Offringa\inst{9}
\and 
L.~E.~Porter\inst{4}
\and
P.~Procopio\inst{15}
\and 
L.~Staveley-Smith\inst{11}
\and 
R.~B.~Wayth\inst{3}
\and 
C.~Wu\inst{11}
\and 
Q.~Zheng\inst{16}
}

\institute{Department of Astronomy, Faculty of Mathematics, University of Belgrade, Studentski trg 16, 11000 Belgrade, Serbia, \email{\emaila}
\and
Western Sydney University, Locked Bag 1797, Penrith South DC, NSW, Australia
\and
International Centre for Radio Astronomy Research, Curtin University, GPO Box U1987, Perth WA 6845, Australia
\and
Space Science Center, Department of Earth and Space Sciences, Morehead State University, 235 Martindale Drive, Morehead, KY 40351, USA
\and
Max-Planck-Institut f\"{u}r extraterrestrische Physik, Giessenbachstra\ss e, 85748 Garching, Germany
\and
School of Physics, The University of New South Wales, Sydney 2052, Australia
\and
School of Physical Sciences, The University of Adelaide, Adelaide 5005, Australia
\and 
University of Technology Sydney, 15 Broadway, Ultimo, NSW 2007, Australia
\and
Netherlands Institute for Radio Astronomy (ASTRON), PO Bus 2, 7990AA Dwingeloo, The Netherlands
\and
Raman Research Institute, Bangalore 560080, India
\and
International Centre for Radio Astronomy Research (ICRAR), M468, University of Western Australia, Crawley, WA 6009, Australia
\and
Centre for Astrophysics Research, School of Physics, Astronomy and Mathematics, University of Hertfordshire, College Lane, Hatfield AL10 9AB, UK
\and
CSIRO Astronomy and Space Science, Marsfield, NSW 1710, Australia
\and
Research School of Astronomy and Astrophysics, Australian National University, Canberra, ACT 2611, Australia
\and
School of Physics, The University of Melbourne, Parkville, VIC 3010, Australia
\and
Shanghai Astronomical Observatory, 80 Nandan Rd, Xuhui Qu, Shanghai Shi, China, 200000
\and
Isaac Newton Institute of Chile, Yugoslavia Branch
}

\abstract{}{In this paper we discuss the radio continuum and X-ray properties of the  so-far poorly studied Galactic supernova remnant (SNR) \snrg.}{We present the radio spectral energy distribution (SED) of the Galactic SNR \snrg\ obtained with the Murchison Widefield Array (MWA). Combining these new observations with the surveys at other radio continuum frequencies, we discuss the integrated radio continuum spectrum of this particular remnant. We have also analyzed an archival \textit{XMM-Newton} observation, which represents the first detection of X-ray emission from this remnant.}{The SNR SED is very well explained by a simple power-law relation. The synchrotron radio spectral index of \snrg\, is estimated to be 0.42$\pm$0.03 
and the integrated flux density at 1~GHz to be around 2.7~Jy. Furthermore, we propose that the identified point radio source, located centrally inside the SNR shell, is most probably a compact remnant of the supernova explosion. The shell-like X-ray morphology of \snrg\ as revealed by {\it XMM-Newton} broadly matches the spatial distribution of the radio emission, where the radio-bright eastern and western rims are also readily detected in the X-ray while the radio-weak northern and southern rims are weak or absent in the X-ray. Extracted MOS1+MOS2+PN spectra from the whole SNR as well as the north, east, and west rims of the SNR are fit successfully with an optically thin thermal plasma model in collisional ionization equilibrium with a column density $N_{\rm{H}}\sim0.80\times10^{22}$ cm$^{-2}$ and fitted temperatures spanning the range $kT\sim0.14-0.23$ keV for all of the regions. The derived electron number densities $n_{\rm{e}}$ for the whole SNR and the rims are also roughly comparable (ranging from $\sim0.20f^{-1/2}$ cm$^{-3}$ to $\sim0.40f^{-1/2}$ cm$^{-3}$, where $f$ is the volume filling factor). We also estimate the swept-up mass of the X-ray emitting plasma associated with \snrg\ to be $\sim46f^{-1/2}$ $M$$_{\odot}$.}{}

\keywords{ISM: individual objects: \snrg -- ISM: supernova remnants -- Radio continuum: ISM -- Radiation mechanisms: general}

\maketitle

\section{Introduction}

Supernova remnants (SNRs) are collisionally ionized emission nebulae that are formed soon after a supernova explosion. In fact, their evolution is closely related to a particular 
type of collisionless shock wave that is actually formed ahead of the ejected material \citep[see][]{V12}. It turns out that SNRs strongly influence the interstellar medium 
through which they expand, and vice versa; furthermore, the ambient interstellar matter (ISM) plays a dominant role in dictating the evolution of an SNR. 

These remarkable objects are believed to be primarily responsible for the production of the Galactic cosmic rays, that is, the high-energy, ultrarelativistic charged particles 
up to $\sim$10$^{15}$\,eV \citep[see, e.g.,][]{Bella,Bellb,BO78}. Of course, one should always bear in mind that such a collisionless shock formation, particle acceleration, and magnetic field amplification 
are all coupled processes that we still do not fully understand \citep[see, e.\!\! g.,][]{Hetal12}. In fact, we still do not fully understand all the processes, like various microinstabilities, 
which actually trigger collisionless shock formation.

The integrated radio continuum spectral energy distribution (SED) of SNRs are generally very well represented by a simple power law, reflecting the pure (non-thermal) synchrotron radiation 
from an SNR shell: 
\begin{equation}
S_{\nu}\propto\nu^{-\alpha},
\end{equation} where $S_{\nu}$ is the spatially integrated flux density and $\alpha$ is the radio spectral index. In that sense, astronomical observations of synchrotron emission, 
predominantly caused by the cosmic ray electrons accelerated via the diffusive shock acceleration (DSA) process, enable us to actually test and constrain all the known theoretical 
models that are believed to describe the physics behind the observed radiation from SNRs \citep[see][for a review]{U14}. It is also easily verified that for a standard value 
of the mean Galactic magnetic field (several $\mu\mathrm{G}$), gigaelectron volt cosmic ray electrons are mostly responsible for the observed synchrotron emission at radio, and 
teraelectron volt electrons at X-ray frequencies \citep{CTA19}.

The mean value of the radio spectral indices, for some $300+$ known Galactic and $\sim$60 Large Magellanic Cloud \citep{2017ApJS..230....2B} SNRs, is around 0.5, which is in a good 
accordance with DSA theory \citep[see, e.g.,][]{Hetal12,GLORIA}. However, our present knowledge of the SNR radio SED's as well as of their overall shape 
is far from being precise, at least for the majority of known Galactic and Magellanic Cloud SNRs. It is therefore very important to obtain more reliable radio spectra, 
so that the analysis of the integrated continuum of SNRs could distinguish between various different theoretical models, particularly in the low-frequency domain.

After a brief review of the current knowledge on the Galactic SNR \snrg, we present the low-frequency ($80-300\ \mathrm{MHz}$) radio observations of this remnant, 
obtained with the Murchison Widefield Array \citep[][see Sects.~2 and 3]{L09,TIN13}. In Sect.~4, we will focus on the analysis of the integrated continuum spectrum 
and morphology of this remnant. In addition, in Sect.~5 we report the first detection of {\snrg} in X-rays and analyse its X-ray spectrum, based on the archived \textit{XMM-Newton} 
observations. Finally, in Sect.~6 we make a summary of our results.

\section{Previous studies of SNR \snrg}
\label{PreviousSNRStudiesSection}
The poorly studied Galactic SNR \snrg\ was originally identified based on the radio continuum surveys of the Galactic plane with the Effelsberg 100-m telescope at $\lambda=11$~cm 
($\nu=2.7$~GHz) and $\lambda=21$~cm ($\nu=1.4$~GHz) \citep{Reich88}. Morphologically, it is an S-type remnant with an asymmetric shell and with an average angular radius of around $10\arcmin$. 
Based on these early radio continuum observations, the radio spectral index of this remnant is found to be around $0.4$ and flux density at 1~GHz around 3.3~Jy \citep{Reich88}. 
Based on a revised radio surface brightness to diameter ($\Sigma-D$) relation, \citet{Pavlov013,Pavlov014} estimated the diameter of the SNR \snrg\ and its distance, based on the method 
of orthogonal fitting, to be around 31.2~pc and 5.4$\pm$2.8~kpc (average fractional error is $0.52$), respectively. In addition, median distance, based on the so-called probability density 
function (PDF) method \citep{Vukotic14} is 5.1$\pm$2.2~kpc (average fractional error is $0.43$).

\snrg\ has been, so far, only observed at radio frequencies and has not been extensively analyzed in the current literature. \cite{HY09} included this remnant in the 
Very Large Array (VLA) 1720~MHz OH maser survey. No detection of interaction between radiative-phase shocks and molecular gas is reported, which is not unusual as the SNR is over 3\D\ 
away from the bulk of the Galactic molecular clouds along the central plane. 

We emphasize that the SNR \snrg\ is also detected in the 1.4~GHz National Radio Astronomy Observatory (NRAO) VLA Sky Survey \citep[NVSS;][]{Condon98} 
as well as in the Tata Institute of Fundamental Research (TIFR) alternative data release 1 (ADR1) Giant Metrewave Radio Telescope (GMRT) 150~MHz Sky Survey\footnote{The GMRT 
is run by the National Centre for Radio Astrophysics of the TIFR.} \citep[TGSS ADR1;][]{INT17}. In addition, this remnant is also identified in the Parkes Radio Observatory 
- Massachusetts Institute of Technology (MIT) - NRAO 4850~MHz Southern Sky Survey \citep[PMN;][]{PMN93}. Unfortunately, because of the missing short spacings within the interferometer 
array configuration, we cannot extract precise flux values for an analysis of the G5.9+3.1 SED using these measurements.

We also note that a (isolated single-Gaussian) point source within the boundaries of \snrg\ has already been detected and listed in the TGSS ADR1 150~MHz survey source catalog 
\citep[J174711.9-221741][]{INT17} at approximately the same apparent location at which we find the central radio source inside the shell of the SNR 
($S_{\mathrm{total}}(150\ \mathrm{MHz})=(263.0\pm28.8)\ \mathrm{mJy}$, $S_{\mathrm{peak}}(150 \mathrm{MHz})=(259.0\pm26.8)\ \mathrm{mJy}$). In addition, 
this point radio source is also listed in the NVSS survey catalog at 1.4~GHz as NVSS J174711-221743 and with $S_{(1.4\ \mathrm{GHz})}=(28.7\pm1.0)\ \mathrm{mJy}$ \citep{Condon98}. 

The SNR \snrg\ has not been detected by the \textit{Fermi} Large Area Telescope (LAT) team \citep{Acero2016}, as well as by any other $\gamma$-ray observations \citep[see][]{HESS18}. 
Also, our extensive search at IR and optical frequencies \citep{stupar08} did not reveal any detection despite \snrg\  being positioned more than 3\D\ above the Galactic 
plane where confusion is expected to be significantly reduced. Finally, \snrg\ was not detected by the Infrared Astronomical Satellite (IRAS) survey of Galactic SNRs \citep{Saken92}, 
nor in the \textit{Spitzer} surveys \citep{Reach06,Andi11}.

\section{Murchison Widefield Array photometry}

In this section we present the aperture photometry method we applied to Murchison Widefield Array (MWA) total intensity maps of the \snrg. The region containing \snrg\ was observed with the MWA as a part of the GaLactic 
and Extragalactic All-sky Murchison Widefield Array (GLEAM) survey~\citep{GLEAM15}. 
Following a similar strategy used to produce the extragalactic radio images in \cite{HW17}, but with improvements in the imaging and deconvolution strategy for the 
confused and complex regions of the Galactic plane, \cite{HW19} produce GLEAM images over the Galactic longitude ranges $345^{\circ} < l < 60^{\circ}$ and $180^{\circ} < l < 240^{\circ}$, 
which covers \snrg. \citet{HW19} also provide flux density measurements for SNR and candidate SNRs within this region; a brief summary of the measurement method is provided here.

Through manual inspection, the region containing the SNR is marked out by the observer. An estimate of the background is made using the average flux density of the surrounding region, 
excluding regions with contaminating structures such as background radio galaxies. This background is subtracted from the total flux density of the area within the SNR region, and then 
the total flux density of the SNR is calculated.

GLEAM provides $20 \times 7.68$\,MHz-bandwidth channels between 72 and 231~MHz. In these images, the object is resolved and clearly distinct from the relatively complicated environment. 
However, at frequencies above 162~MHz, the signal-to-noise ratio becomes somewhat low and the flux density measurements less accurate. For \snrg, we therefore use the wideband 
GLEAM measurements of 72--103~MHz, 103--134~MHz, 139--170~MHz, and 170--231~MHz, for improved signal-to-noise ratio.

\begin{figure*}
\centering
\includegraphics[width=1\linewidth, trim=0 180 0 170]{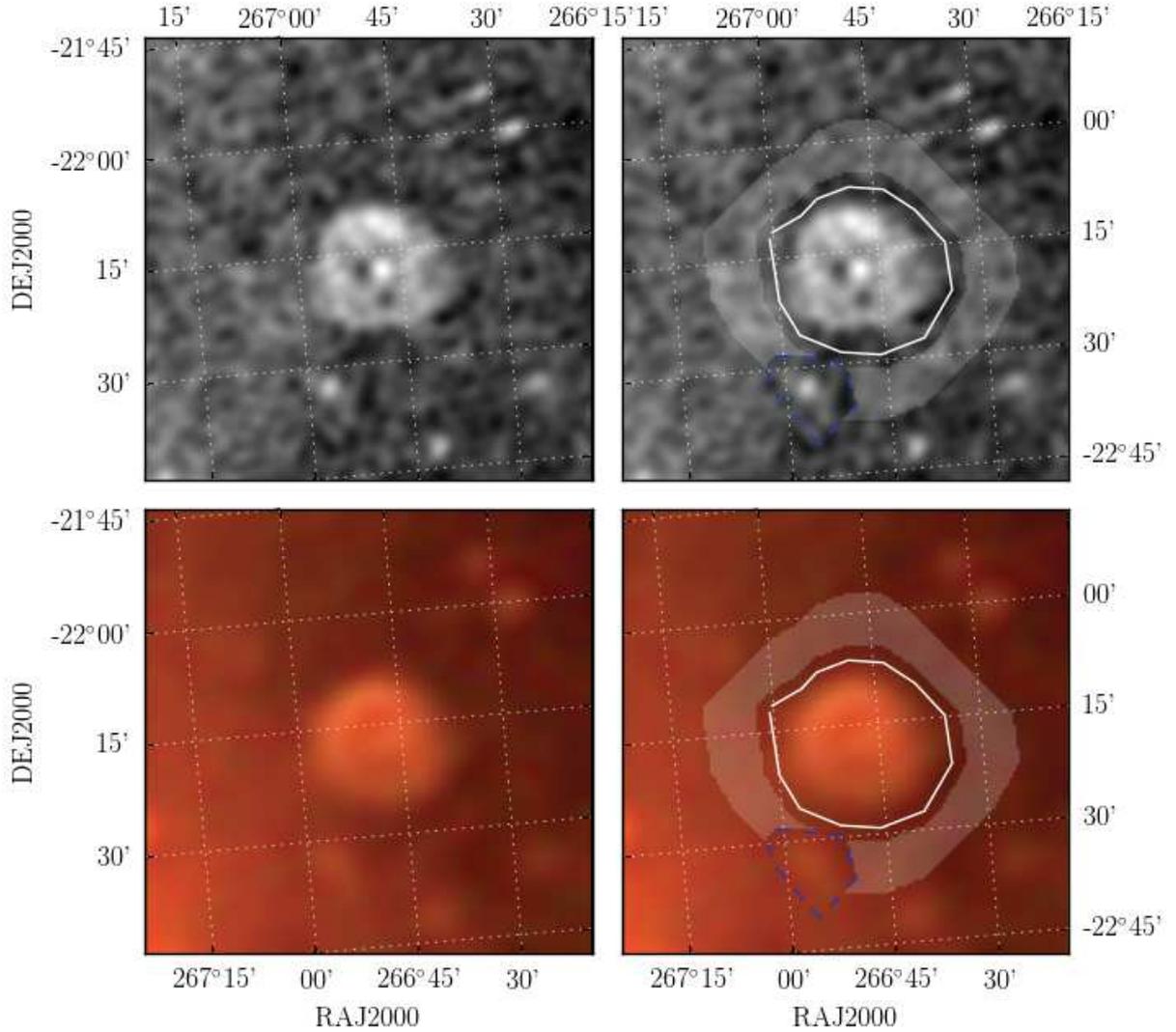}
\caption{GLEAM images of \snrg; the top two panels show the wideband image taken over 170--231~MHz, while the lower two panels show an RGB image comprised of R$=72$--$103$~MHz, 
G$=103$--$134$~MHz, B$=139$--$170$~MHz. The full width at half maximums (FWHMs) of the point spread functions (PSFs) of the images are, respectively: 2.4$\arcmin$, 5.2$\arcmin$, 3.9$\arcmin,$ and 2.9$\arcmin$. The left panels 
show the images without any annotations. The right panels show annotations indicating the GLEAM photometry measurement: the white lines show the region drawn by the observer 
that contains the SNR; the gray shaded region shows the area used to calculate the background; the blue dashed lines show a region excluded from the background measurement 
for containing contaminating background radio galaxies.}
\label{fig:G5.9+3.1_image}
\end{figure*}

Figure~\ref{fig:G5.9+3.1_image} shows the GLEAM measurements in these bands, with the highest resolution 170--231\,MHz image in the top two panels and the three remaining 
bands as an red-green-blue (RGB) image in the lower two panels. The SNR is quite circularly symmetric, with a diameter of $18.5\arcmin$.

\begin{table}[t]
\caption{MWA, Effelsberg, and RATAN-600 integrated flux density measurements of SNR \snrg.}
\centering
\begin{tabular}{@{}llc}
\hline
Frequency -- $\nu$            & Telescope & $S_{\nu}^{\mathrm{SNR}}$  \\
$(\mathrm{MHz})$ &           & $(\mathrm{Jy})$           \\
\hline
072-103 (87.5)  & MWA        & $8.5\pm0.9$  \\
103-134 (118.5) & MWA        & $7.3\pm0.7$  \\
139-170 (154.5) & MWA        & $5.9\pm0.6$  \\
170-231 (200.5) & MWA        & $4.3\pm0.4$  \\
960             & RATAN-600  & $4.0\pm0.4$  \\
1408            & Effelsberg & $2.4\pm0.2$  \\
2695            & Effelsberg & $1.8\pm0.3$  \\
3900            & RATAN-600  & $1.3\pm0.2$  \\
\hline
\label{tbl:mwaphoto}
\end{tabular}
\end{table}

\section{Integrated radio continuum spectrum of the supernova remnant \snrg}
\label{IntegratedRadioContinuumSpectrumSection}

\begin{figure*}[t]
\centering
\includegraphics[width=0.8\linewidth]{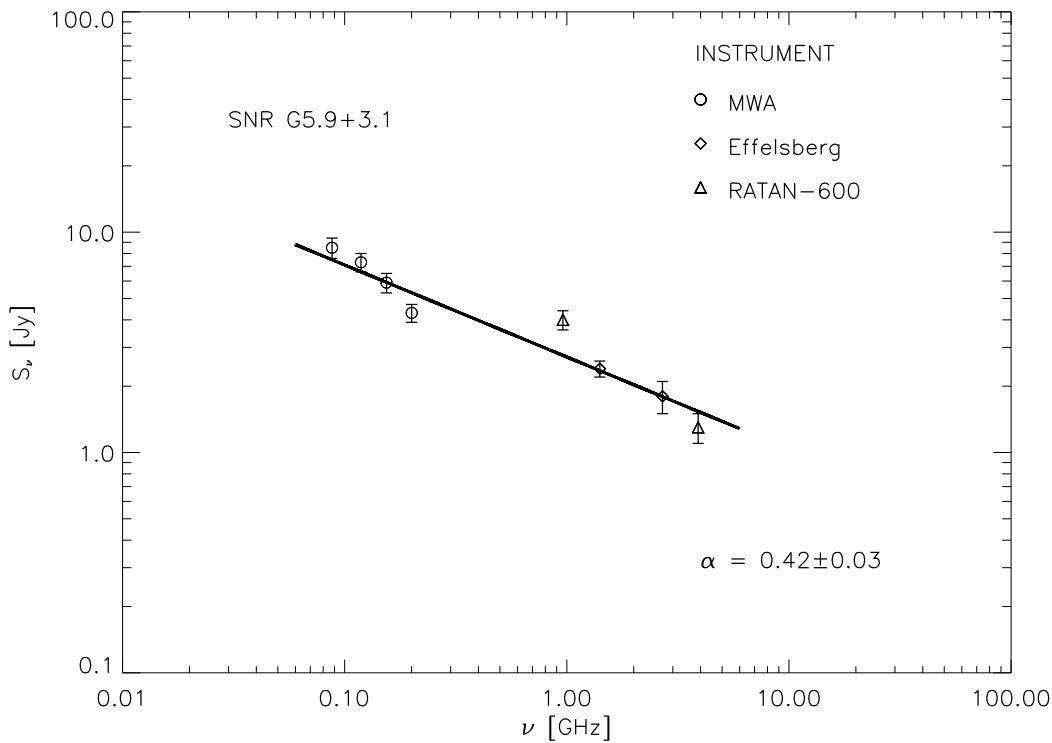}
\caption{Weighted least-squares fit of the presently known integrated radio continuum spectrum of the SNR \snrg\ using the simple power-law model (Eq.~1) 
for pure synchrotron emission (solid line). Circles correspond to the MWA data, diamond symbols indicate Effelsberg data, and triangles depict RATAN observations.}
\label{SEDsnr}
\end{figure*}

Combining particular MWA wideband low-frequency observations with the Effelsberg 100-m Galactic plane survey at 1408~MHz and 2695~MHz \citep{EFELSI,EFELSII} 
as well as with the Academy of Sciences Radio Telescope - 600 (RATAN-600) Galactic plane survey at 960~MHz and 3900~MHz \citep{RATAN96,RATAN}, we present the SED of the 
SNR \snrg\ (see Fig.~\ref{SEDsnr} and Table~\ref{tbl:mwaphoto}). The integrated flux densities for Effelsberg data are estimated from the publicly available observations at 
Max Planck Institute for Radio Astronomy (MPIfR) Survey Sampler\footnote{\url{http://www3.mpifr-bonn.mpg.de/survey.html}.}, and for the RATAN-600, 
integrated flux densities are taken from \citet{RATAN96,RATAN}.

Because of the steep radio spectral index ($\alpha=0.99\pm0.05$ from TGSS and NVSS observations) and small values of the flux densities for the point radio source, 
which lay in the center of the SNR shell, we can actually use the MWA, Effelsberg, and RATAN-600 integrated flux densities assuming that the contribution of the central 
radio source is negligible.

No obvious SED turn-off is detected at the lowest observed continuum frequencies (see Fig.~\ref{SEDsnr}). In a similar fashion, a curvature at higher frequencies (up to 3.9~GHz) 
is also missing. The SNR SED is very well explained by a simple power-law relation.

\begin{figure*}[t]
\centering
\includegraphics[width=0.8\linewidth]{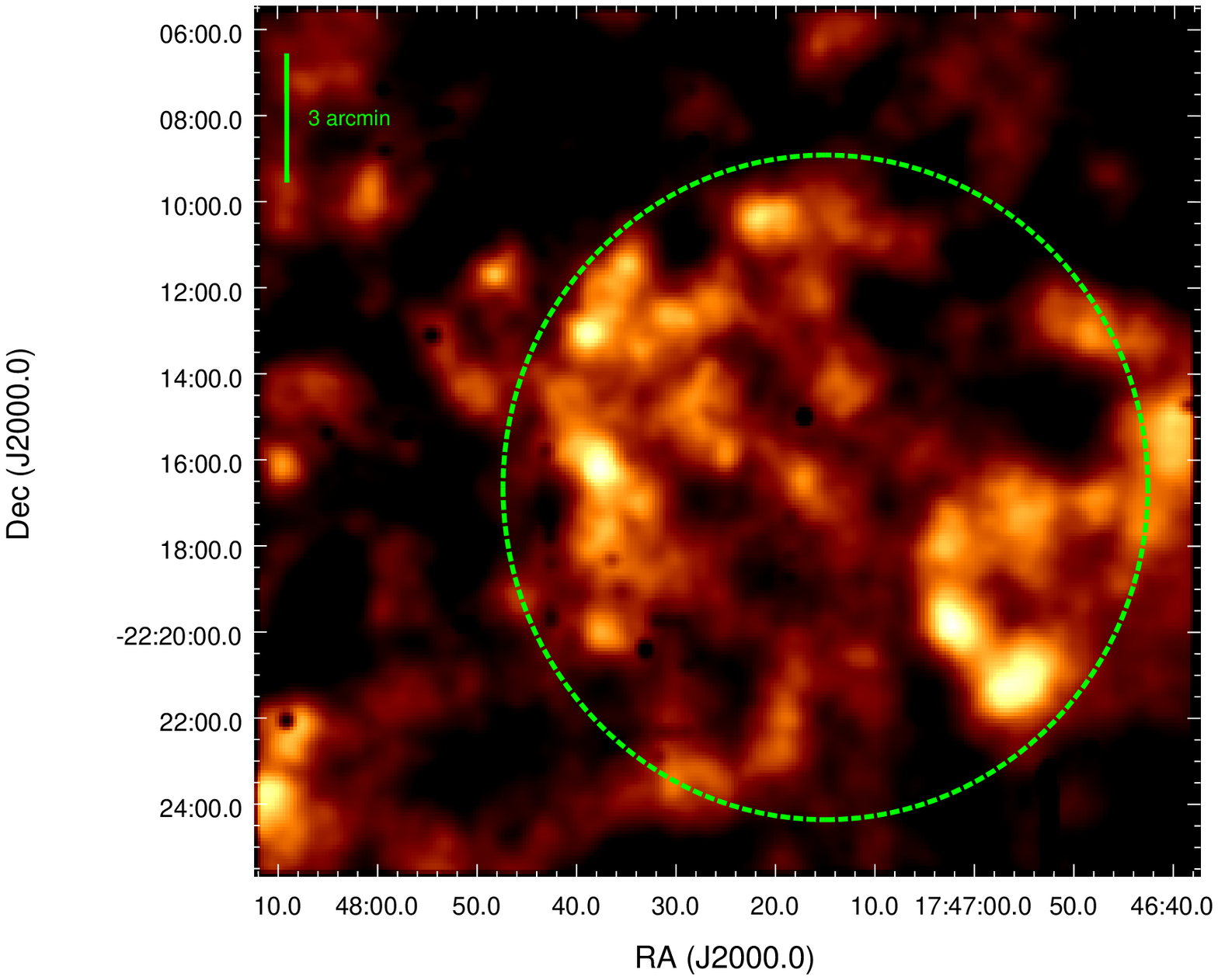}
\caption{Adaptively smoothed exposure-corrected combined EPIC image of G5.9+3.1. The image depicts emission detected over the energy range 0.4 keV 
to 1.0 keV and the pixel range of the emission is 3 to 7.87 counts per second per square degree. The dashed green ellipse is centered on RA (J2000.0) 17$^\mathrm{h}$ 47$^\mathrm{m}$ 15.0$^\mathrm{s}$, Dec (J2000.0) $-22^{\circ}$ 16$\arcmin$ 38.5$\arcsec$ and has 
major and minor axes of 449.5$\arcsec$ and 463.4$\arcsec$, respectively. This ellipse corresponds to the region of extraction
of the source spectrum for the whole SNR (see Sect. \ref{XMMObservationsSection} and Table 
\ref{RegionPropertiesTable}). We note the clear rim-brightened structure seen toward the eastern and southwestern rims of G5.9+3.1.
Fainter emission is seen toward the northern rim and interior of the SNR. Emission from discrete sources has
been flagged and excised in the creation of this image: voids of emission seen in this image correspond to
artifacts from these excisions.}
\label{G5_XMMImage}
\end{figure*}

In Fig.~\ref{SEDsnr}, we present the weighted least-squares fit of the integrated radio continuum (over the the whole observable frequency range) of the SNR \snrg\ using the 
simple power-law model for pure synchrotron emission (Eq.~1). Circles correspond to the MWA data, diamond symbols indicate Effelsberg data and triangles depict RATAN observations. 
The best fitting radio synchrotron spectral index $\alpha$ is around 0.42 and the flux density at 1~GHz is estimated to be around 2.7~Jy. 

The weighted least-squares fit is estimated with the \verb"MPFIT"\footnote{http://purl.com/net/mpfit.} \citep{MPFIT} package written in \verb"IDL", with starting values estimated 
from the data. We note that \verb"MPFIT" provides estimates of the 1$\sigma$ uncertainties for each parameter. 

Furthermore, because the spectral index of \snrg\ is less than the value of 0.5 predicted by test-particle DSA, the so-called equipartition calculation, derived for $\alpha\geq0.5$, 
cannot be used for magnetic field estimation for this SNR \citep{AUAPV12,AUVPV13,UPA18}. Bearing in mind the calculated parameter uncertainty, spectral indices less than 0.5 can be explained 
by a significant contribution of the stochastic Fermi acceleration \citep[see][]{OW99,OS93,SF89,onic13}. 

This central point radio source is coincident ($\sim$$20\arcsec$) with the position of the long-period variable (LPV) star known in the Optical Gravitational Lensing Experiment 
Galactic Bulge LPV star catalogue (OGLE BLG-LPV) as \object{OGLE BLG-LPV 38416} \citep{OGLE13}. However, the non-thermal radio continuum of this point source is not in a good accordance with the optically thick thermal free-free emission from a post-shock partially ionized layer in a stellar atmosphere, or with thermal emission from the stellar photosphere, nor with non-thermal radio emission associated with stellar flare activity \citep[see][]{Estalella83,LPV,RM97,RC07}. Furthermore, the nearest known pulsar PSR+J1745-2229 \citep{Hobbs04} is about $20'$ from the point source.

Additionally, in contrast to the radio spectral indices of SNRs (usually between 0.3 and 0.8), pulsar wind nebulae (PWNe) have spectral indices in the range of around 
$0-0.3$ \citep[][and references therein]{PWN12}. On the other hand, above 100~MHz the spectra of the majority of detected pulsars can be described just by a simple power law with 
the average value of spectral index around 1.8 \citep{PULSARI}. \citet{PULSARII} have found that spectral indices of 228 pulsars from their analysis are distributed in the range 
from 0.46 to 4.84. In that sense, we propose that this central object is possibly the compact remnant of the supernova explosion that gave rise to \snrg.

\section{\textit{XMM-Newton} observations of {\snrg}}
\label{XMMObservationsSection}

{The SNR \snrg} was the subject of a pointed observation made with the three European Photon 
Imaging Cameras (EPIC) Metal Oxide Semi-conductor (MOS) 1, MOS2, and p-n (PN) aboard the \textit{XMM-Newton} 
Observatory on 1~March~2006 (PI: R. Bandiera; Obs.~ID 0553110401). The archival 
datasets from the observations were downloaded from the \textit{XMM-Newton} Science 
Archive and analyzed using standard tools in the High Energy Astrophysics Software (HEASOFT) Package (Version~6.22.1) 
and the Science Analysis Software (SAS) software package (Version~16.1.0).
The SAS tools \verb"epchain" and \verb"emchain" were used to apply standard processing 
tools to the PN, MOS1, and MOS2 datasets while the tools \verb"pn-filter" and 
\verb"mos-filter" 
were used to filter the datasets for background flaring activity. After processing, 
the effective exposure times of the PN, MOS1, and MOS2 cameras were 6281~seconds, 
11040~seconds, and 11080~seconds, respectively. Combined 
EPIC (PN+MOS1+MOS2) exposure-corrected and adaptively smoothed images of \snrg\ were 
created using the Extended Source Analysis Software package \citep{SK11}.

\begin{figure*}[t]
\centering
\includegraphics[width=0.8\linewidth]{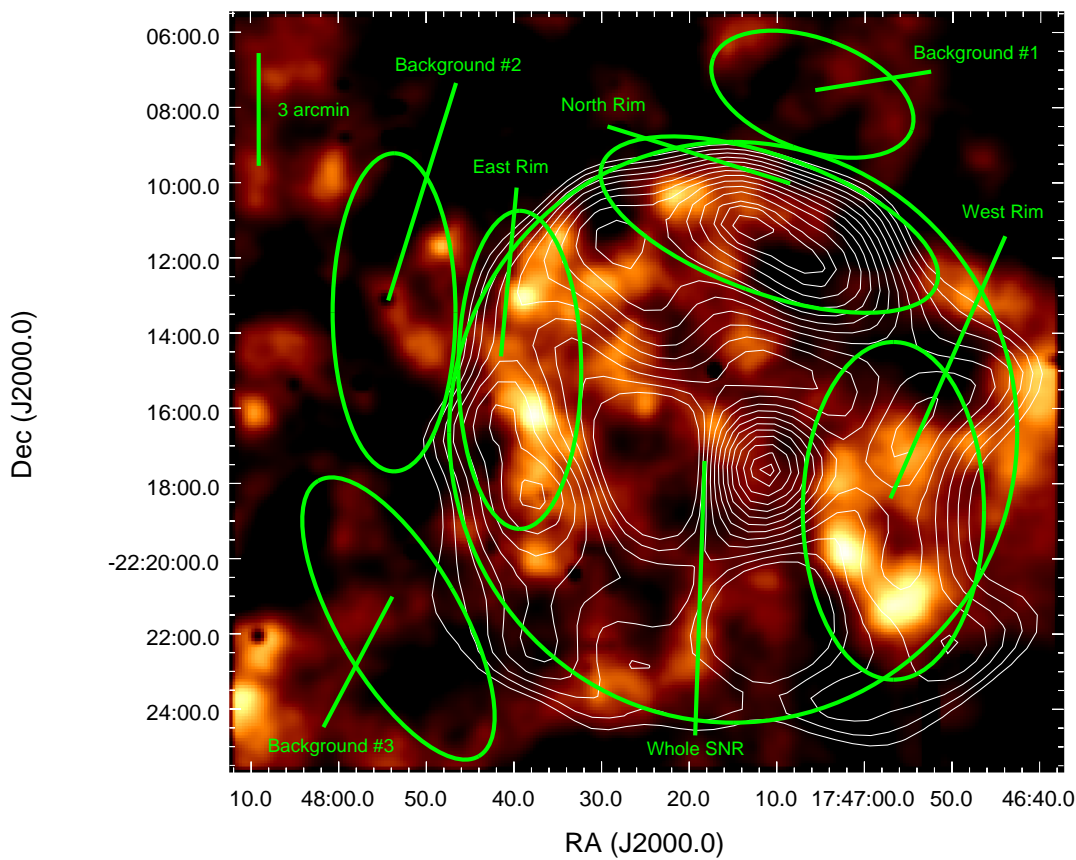}
\caption{Same as Fig. \ref{G5_XMMImage} but with contours overlaid that depict radio emission detected by the MWA and 
regions of spectral extraction indicated and labeled. The contour levels correspond to 0.03, 0.04, 0.05, 0.06, 0.07, 0.08,
0.09, 0.10, 0.11, 0.12, 0.13, 0.14, 0.15, 0.155, and 0.16 Jy/beam. The X-ray emission along the eastern and western
rims of \snrg\  appears to lie interior to the radio rims, while X-ray and radio emission from the northern and southern rims
are weak or absent. The source regions of spectral extraction for
the northern, western, and eastern rims along with the entire SNR are indicated, as well as the accompanying regions of 
background spectral extraction.} 
\label{G5_XMMImage_RadioContours_Xray}
\end{figure*}
\par
In Fig. \ref{G5_XMMImage} we present an exposure-corrected and adaptively smoothed combined EPIC image of {\snrg}. 
The image depicts emission detected
over the energy range from 0.4 keV to 1.0 keV. The dashed green ellipse
is centered on RA (J2000.0) 17$^\mathrm{h}$ 47$^\mathrm{m}$ 15.0$^\mathrm{s}$, Dec (J2000.0) $-$22$^{\circ}$ 16$\arcmin$ 38.5$\arcsec$ and has 
major and minor axes of 449.5$\arcsec$ and 463.4$\arcsec$, respectively. This ellipse corresponds to the region of extraction
of the source spectrum for the entirety of \snrg\, and the spectral properties of this region -- along with smaller 
regions of the SNR -- are discussed later in this section. A shell-like structure is clearly visible: the eastern rim (toward the left of the image) is particularly noticeable while the southwestern
rim is readily detected as well. Fainter emission is detected along the northern rim and the interior of \snrg,\ while the southern portion of the SNR is not detected in the
X-ray. We believe this result is the first published detection of X-ray emission from {\snrg}: we also note that the aimpoint of the observation (where the angular resolution capabilities
and flux sensitivities of the imaging instruments attain their best values) was placed on the eastern rim of the SNR. Therefore, significant degradation in both angular
resolution and flux sensitivity with increasing off-axis angle -- particularly in the western direction toward the western edge of the radio shell of \snrg\ -- complicates the
development of a detailed and accurate description of the X-ray morphology of this SNR. From inspection of the available image presented in Fig.
\ref{G5_XMMImage}, we conclude that the X-ray morphology of \snrg\ is best described as more shell-like than center-filled. Additional X-ray observations of this SNR are
needed to probe its X-ray morphology in a more rigorous manner. Finally, we note that the central radio point source 
discussed in detail in Sect. \ref{IntegratedRadioContinuumSpectrumSection} is not detected in our X-ray
image.

\begin{table*}[t]
\begin{center}
\caption{Properties of regions of extraction of source and background spectra for G5.9$+$3.1.}
\label{RegionPropertiesTable}
\begin{tabular}{lccccc} 
\hline
& RA & Dec & Major & Minor & Position \\
& (J2000.0) & (J2000.0) & axis & axis & angle \\
Region & (h m s) & ($^{\circ}$ $\arcmin$ $\arcsec$) & ($\arcsec$) & ($\arcsec$) & ($^{\circ}$) \\
\hline
Whole SNR (Source) & 17 47 15.0 & $-$22 16 38.5 & 449.5 & 463.4 & 0 \\
East rim (Source) & 17 47 39.3 & $-$22 14 59.0 & 97.2 & 253.8 & 0 \\
North rim (Source) & 17 47 10.9 & $-$22 11 07.1 & 109.2 & 281.4 & 70 \\
West rim (Source) & 17 46 56.7 & $-$22 18 43.7 & 143.4 & 269.4 & 180 \\ 
Background region \#1 & 17 47 06.0 & $-$22 07 38.9 & 89.1 & 167.1 & 70 \\
Background region \#2 & 17 47 53.2 & $-$22 21 35.8 & 97.2 & 253.8 & 30 \\
Background region \#3 & 17 47 53.7 & $-$22 13 27.2 & 97.2 & 253.8 & 0 \\
\hline
\end{tabular}
\end{center}
\end{table*}

\begin{table*}[t]
  \begin{center}
    \caption{Summary of fit to extracted {\it XMM-Newton} source and background 
    spectra of G5.9+3.1 Using TBABS$\times$APEC model.}
        \label{SpectralFitParametersTable}
    \begin{tabular}{lcccc} 
   \hline
      Parameter & Whole SNR & East rim & North rim & West rim \\
      \hline
\multicolumn{5}{c}{Source spectrum fit parameters} \\
      \hline
      $N$$_{\rm{H}}$ (10$^{22}$ cm$^{-2}$) & 0.80$\pm$0.04 & 0.75$^{+0.07}_{-0.08}$ & 0.84$^{+0.15}_{-0.09}$ & 0.80$^{+0.10}_{-0.11}$ \\ 
      $kT$ (keV) & 0.15$^{+0.03}_{-0.01}$ & 0.14$^{+0.04}_{-0.03}$ & 0.17$^{+0.08}_{-0.06}$ & 0.23$^{+0.10}_{-0.06}$  \\
      Abundance (Solar) & 1.00 (frozen) & 1.00 (Frozen) & 1.00 (Frozen) & 1.00 (Frozen) \\
      Normalization (cm$^{-5}$) & 4.15$\times$10$^{-2}$ & 5.81$\times$10$^{-3}$ & 1.89$\times$10$^{-3}$ & 2.40$\times$10$^{-3}$  \\
      \hline
\multicolumn{5}{c}{Astrophysical X-ray background spectrum fit parameters} \\
\hline
Constant (PN-SRC) & 1.00 (Frozen) & 1.00 (Frozen) & 1.00 (Frozen) & 1.00 (Frozen) \\
Constant (PN-BKG) & 1.00 (Frozen) & 1.00 (Frozen) & 1.00 (Frozen) & 1.00 (Frozen) \\
Constant (MOS1-SRC) & 0.85 (Frozen) & 0.85$\pm$0.04 & 0.85 (Frozen) & 0.83$\pm$0.05 \\
Constant (MOS1-BKG) & 0.85 (Frozen) & 0.85$\pm$0.04 & 0.85 (Frozen) & 0.83$\pm$0.05 \\
Constant (MOS2-SRC) & 0.85 (Frozen) & 0.83$\pm$0.04 & -- & -- \\
Constant (MOS2-BKG) & 0.85 (Frozen) & 0.83$\pm$0.04 & -- & -- \\
\hline
$kT$$_{\rm{Halo}}$ (keV) & 0.30$^{+0.02}_{-0.01}$ & 0.29$^{+0.02}_{-0.01}$ & 0.28$^{+0.02}_{-0.04}$ & 0.28$^{+0.03}_{-0.04}$  \\
Abundance (Solar) & 1.00 (Frozen) & 1.00 (Frozen) & 1.00 (Frozen) & 1.00 (Frozen) \\
Normalization$_{\rm{Halo}}$ (cm$^{-5}$) & 1.35$\times$10$^{-2}$ & 1.58$\times$10$^{-3}$ & 2.62$\times$10$^{-3}$ & 2.89$\times$10$^{-3}$  \\ 
$\Gamma$$_{\rm{DXB}}$ & 1.46 (Frozen) & 1.46 (Frozen) & 1.46 (Frozen) & 1.46 (Frozen) \\
Normalization$_{\rm{DXB}}$ & 1.60$\times$10$^{-3}$ & 1.64$\times$10$^{-4}$ & 2.12$\times$10$^{-4}$ & 2.83$\times$10$^{-4}$  \\
\hline
$kT$$_{\rm{LHB}}$ (keV) & 0.10 (Frozen) & 0.10 (frozen) & 0.10 (Frozen) & 0.10 (Frozen) \\
Abundance (Solar) & 1.00 (Frozen) & 1.00 (Frozen) & 1.00 (Frozen) & 1.00 (Frozen) \\
Normalization$_{\rm{LHB}}$ (cm$^{-5}$) & 2.42$\times$10$^{-3}$ & 2.60$\times$10$^{-4}$ & 3.62$\times$10$^{-4}$ & 4.55$\times$10$^{-4}$  \\
\hline
C-Statistic & 5510.72 & 3230.99 & 3614.57 & 3517.77  \\
Degrees of Freedom & 5473 & 4033 & 3675 & 3674  \\
\hline
\multicolumn{5}{l}{{\bf Notes:} All quoted error bounds correspond to the 90\% 
confidence levels. In the case of the APEC model,} \\ 
\multicolumn{5}{l}{the normalization is defined as 
(10$^{14}$/4$\pi$$d$$^2$)$\int n_{\rm e}n_{\rm p}dV$, where $d$ is the distance
to the SNR (in units of centimeters),} \\ 
\multicolumn{5}{l}{$n_{\rm{e}}$ and $n_{\rm{p}}$ are the 
number densities of electrons and protons respectively (in units of cm$^{-3}$), 
and finally} \\
\multicolumn{5}{l}{$\int dV=V$ is the integral over the entire volume (in 
units of cm$^3$). In the case of the power-law model, the}\\
\multicolumn{5}{l}{the normalization is defined as photons keV$^{-1}$ cm$^{-2}$ 
s$^{-1}$ at 1 keV (see Sect. \ref{XMMObservationsSection}).}
    \end{tabular}
  \end{center}
\end{table*}

In Fig. \ref{G5_XMMImage_RadioContours_Xray} we present the same image given in Fig. \ref{G5_XMMImage} but with contours added to depict radio emission detected by MWA from \snrg. 
Inspection of this figure reveals that the bright X-ray emission from the eastern and western rims of this SNR appears to lie
interior to the radio emission from these rims. We note that the eastern rim of the SNR is the brightest feature of the SNR in both the X-ray and the radio. The northern and 
southern rims of the SNR are weak in radio and essentially absent in the X-ray while little to no X-ray emission is seen
along the southern radio rim. 
\par
Using the data collected by all three EPIC cameras aboard {\it XMM-Newton} during this observation, we conducted a 
spatially-resolved spectroscopic analysis of the 
X-ray emission from {\snrg}. 
This analysis was conducted using the standard X-ray spectroscopic analysis software package \citep[XSPEC;][]{Arnaud96} version 12.10.0c and can be described as follows. 
Appropriate source regions (corresponding to the whole SNR along with the
east rim, the north rim, and the west rim) and background regions 
were selected taking care 
to exclude the point sources. The locations of these regions of spectral extraction are indicated in 
Fig. \ref{G5_XMMImage_RadioContours_Xray} and the properties of these regions -- including central RA (J2000) and 
Dec (J2000) values and dimensions -- are listed in Table \ref{RegionPropertiesTable}. As the background can 
have strong spatial 
variations, the region for extraction of background spectra was selected by averaging over several 
regions across the detector plane. Two spectra (source and background) were extracted per instrument
(PN, MOS1, and MOS2) from each region: the first spectrum was extracted from the event
list of the science observation and the second spectrum was extracted from the filter wheel closed (FWC) data.
The FWC spectra were extracted at the same detector position
as in the science observation because of the strong position dependency
of the instrumental background for the PN, MOS1, and MOS2 cameras. 
The corresponding spectra from the FWC data were subtracted from the 
science spectra of the source and background regions to subtract the 
quiescent particle background component. Finally, in anticipation of a statistical analysis
of the data using C-statistics \citep{C79} the extracted spectra were
all rebinned to a minimum of one count per bin.
\par
The source spectra associated with emission from \snrg\ (either from the whole SNR or from one
of its rims) were fitted with the thermal model known as Astrophysical Plasma Emission Code \citep[APEC;][]{Foster12}: this model describes 
an optically thin thermal plasma in collision ionization equilibrium (CIE). For our purposes, the elemental abundances of 
the plasma were frozen to solar values. Furthermore, the source spectra for each region of interest of \snrg\ and the background spectra from 
each camera (PN, MOS1, and MOS2) were fit simultaneously to constrain the source component as well as
the astrophysical background (AXB) component. Following the 
example given by \citet{KS10}, we used a three-component
model for the AXB: the first component was an unabsorbed thermal component for the local hot bubble
(LHB), while the second component was an absorbed thermal component for the
Galactic halo emission. Both of these components were fit with the 
APEC model again with elemental abundances frozen to solar values. Lastly, the third component was an absorbed 
power law for the non-thermal unresolved extragalactic X-ray background. These three components will be denoted
as the LHB component, the Halo component, and the DXB component, respectively, for the remainder of this paper.   
The full model applied in the spectral fitting can be denoted as CONSTANT$\times$(APEC$_{\rm{LHB}}$ + 
TBABS$\times$(APEC$_{\rm{Halo}}$ + Power Law$_{\rm{DXB}}$) + TBABS$\times$APEC$_{\rm{Source}}$. Here, "CONSTANT" refers to multiplicative constants used to account for variations in calibration between the different EPIC cameras and "TBABS" is the T\"{u}bingen-Boulder interstellar photoelectric absorption model (the interstellar elemental abundances given by \citet{WAMc00} were used in our analysis). The intercalibration constants were determined from the spectrum with the highest statistical quality (west rim), and was fixed to these values for the rest of the spectral fits. The column densities for the source
spectra, the Halo component, and the DXB components were all tied together during the fitting. The temperature
of the LHB component was allowed to vary during the fitting \citep[its fitted value of $kT\sim0.30$ keV is comparable
to the fitted temperature of the component obtained by][]{KS10} while the temperature of the Halo component was frozen to $kT=0.10$ keV \citep[which again is consistent with the results of][]{KS10}. Finally, the spectral
index of the DXB component was fixed to 1.46 \citep{CFG97}. We note that we excluded
the extracted MOS2 spectra for the north rim and the west rim from our spectral
analysis of those particular regions of \snrg\ due to poor signal-to-noise.
\begin{figure*}[t]\vspace{2cm}
\centering
\includegraphics[width=0.35\linewidth, angle=-90]{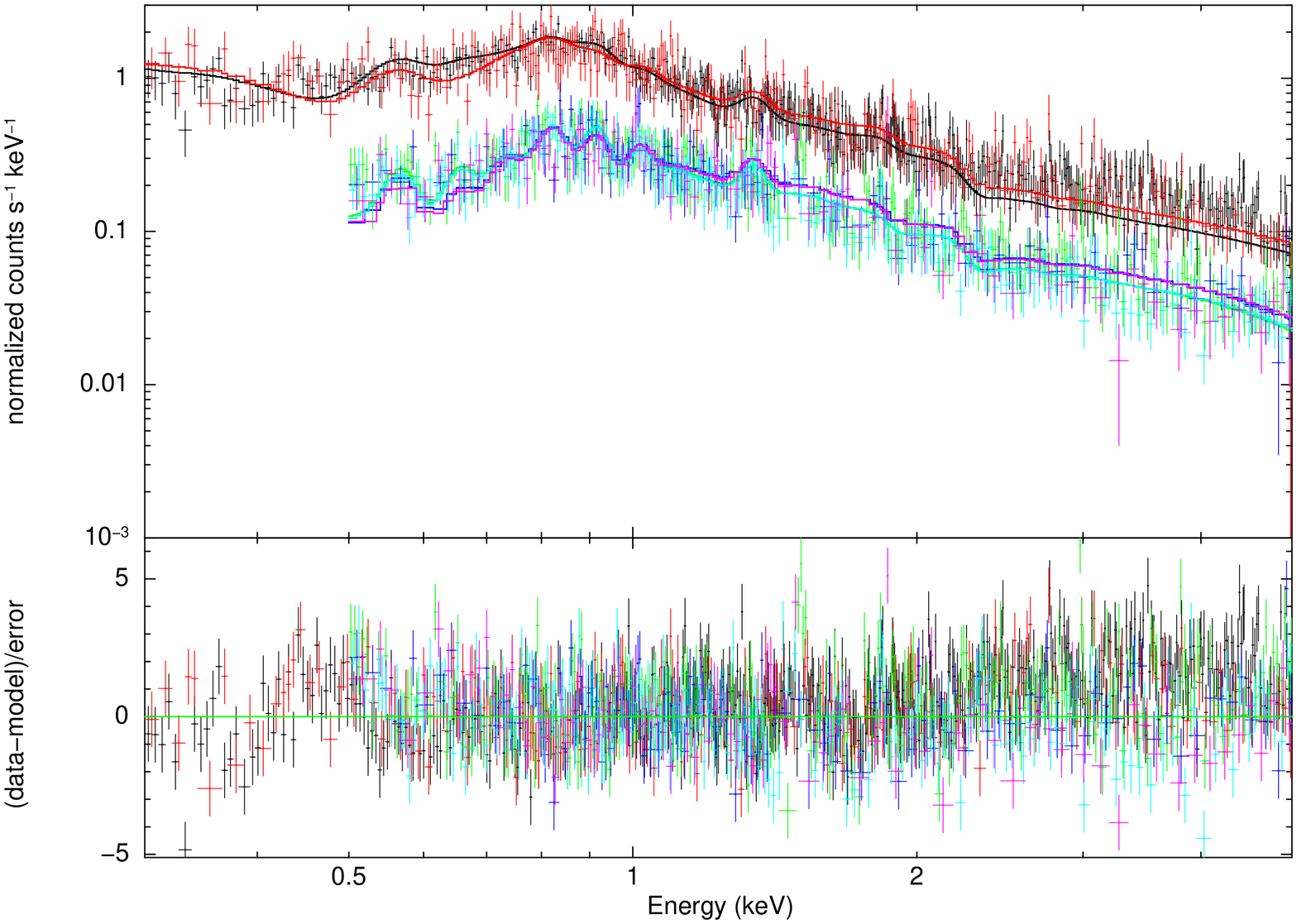}
\includegraphics[width=0.35\linewidth, angle=-90]{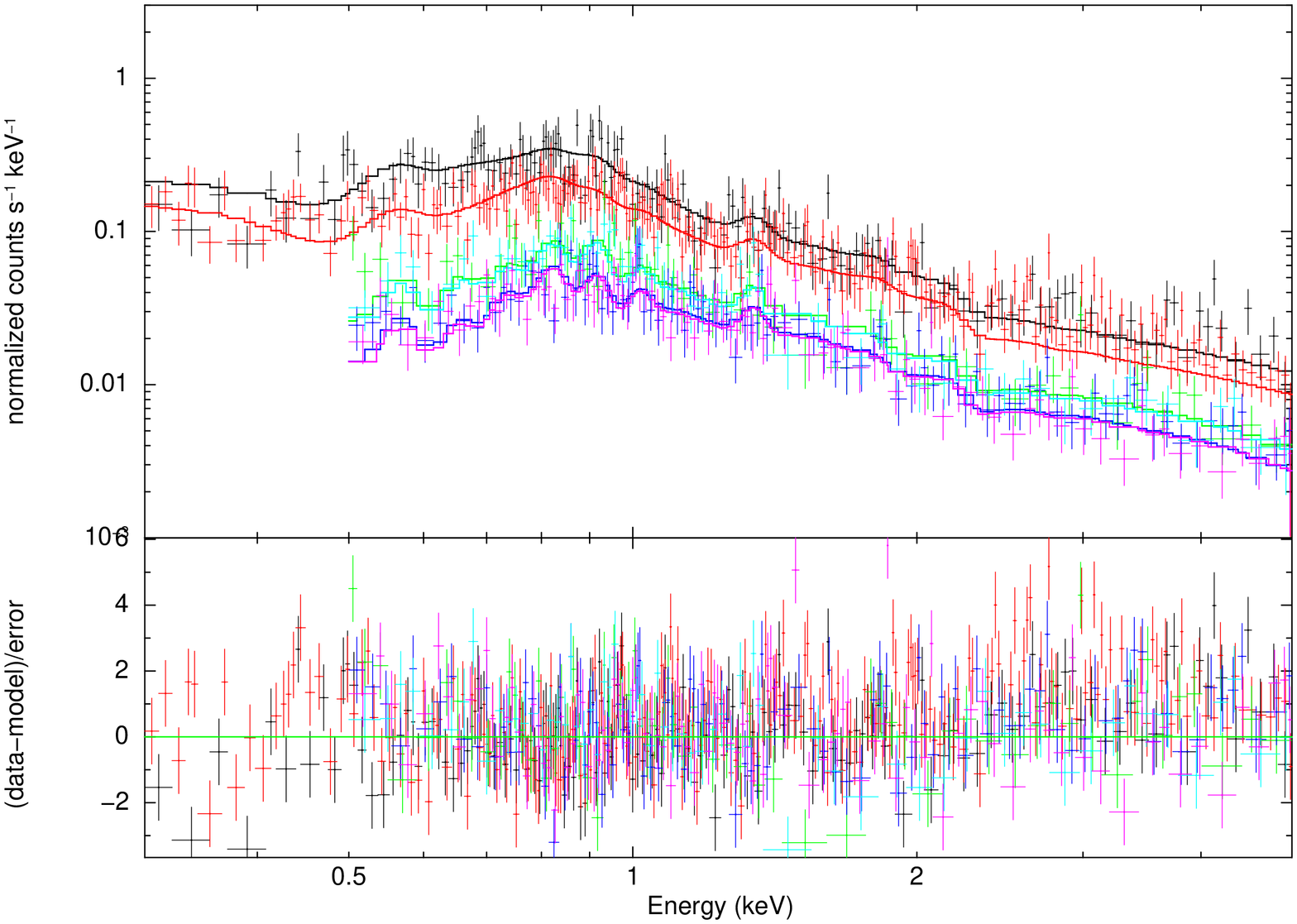}
\includegraphics[width=0.35\linewidth, angle=-90]{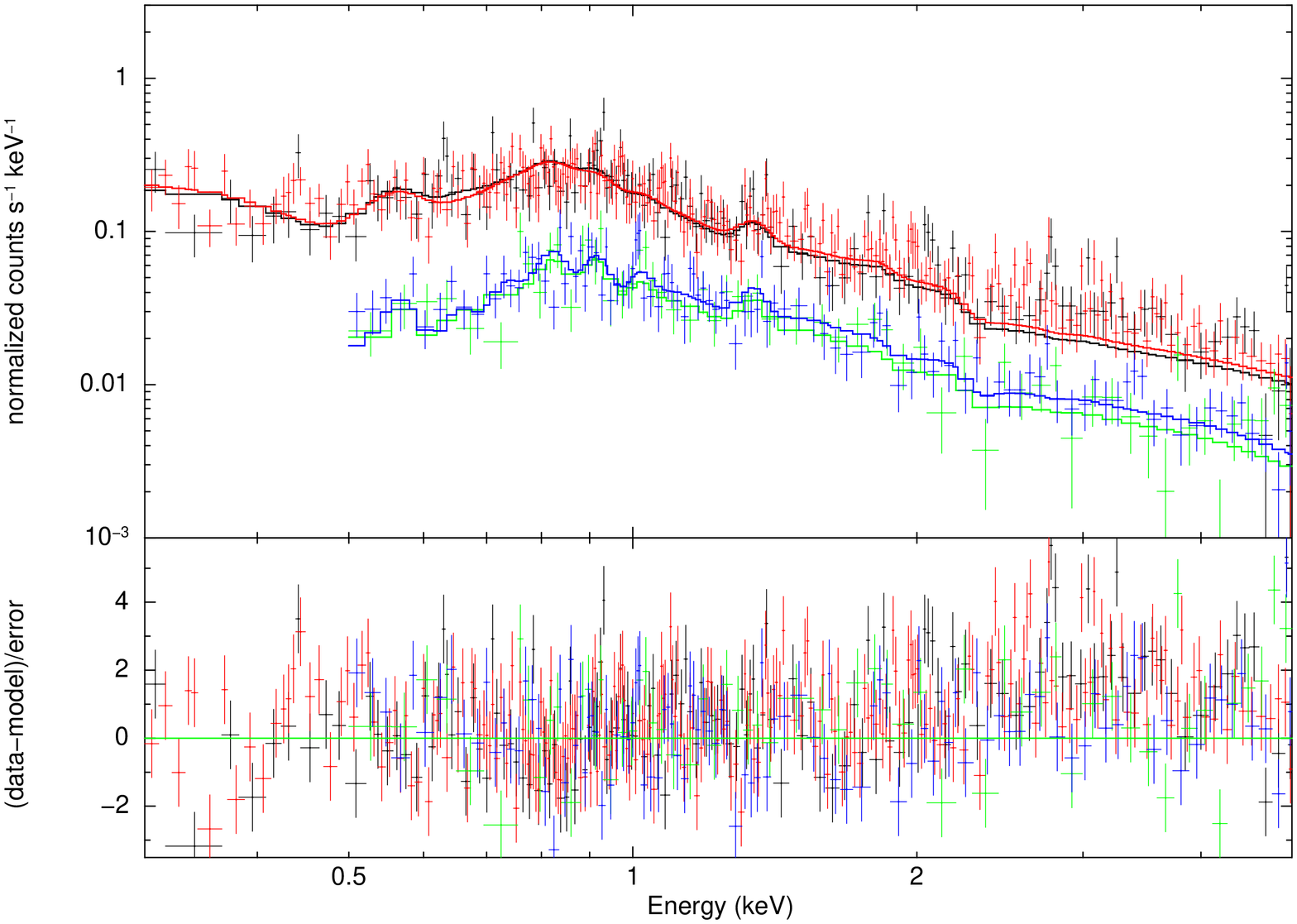}
\includegraphics[width=0.35\linewidth, angle=-90]{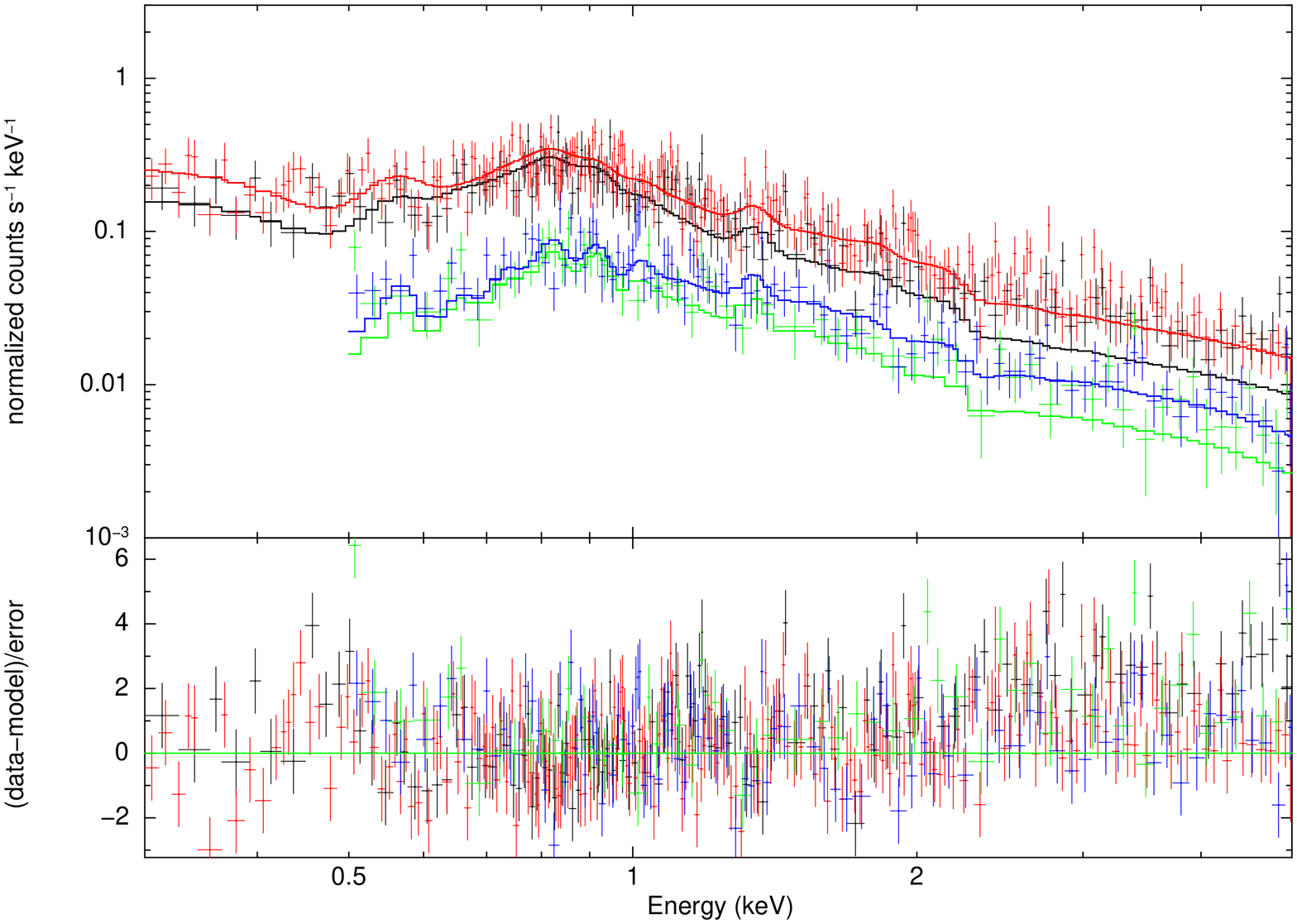}
\caption{Extracted PN, MOS1, and MOS2 spectra of the different regions of G5.9+3.1 as fit with the model and fit 
parameters given in Table \ref{SpectralFitParametersTable}. The residuals to the fits are also shown. In all panels, the
PN source spectrum is shown in black, the PN FWC spectrum is shown in red, the MOS1 source spectrum is shown in
green, the MOS1 FWC spectrum is shown in blue, the MOS2 source spectrum is shown in cyan, and the MOS2 FWC spectrum is shown
in purple. (a -- Upper left) Extracted spectra for the whole SNR. (b -- Upper right) Extracted spectra for the east rim.
(c -- Lower left) Extracted spectra for the north rim. (d -- Lower right) Extracted spectra for the west rim.}
\label{G5_TBABSAPEC_3Rims_Spectra}
\end{figure*}
\par
In Table \ref{SpectralFitParametersTable} we present the results of our spectral fitting
with the model described above. In Fig. \ref{G5_TBABSAPEC_3Rims_Spectra} we present
plots of the extracted spectra for the whole SNR along with the three rims as fit
with the model described above. We have obtained a statistically acceptable fit
to this model based on the interpretation of Cash statistics. The fitted value of the 
column density toward \snrg\ is $N_{\rm{H}}\sim0.80\times10^{22}$ cm$^{-2}$ and the fitted values for the temperatures
of the whole SNR and the three rims are approximately similar, ranging
from $kT\sim0.14$ keV to $kT\sim0.23$ keV. There is significant
overlap in the error bounds on these fitted values: we therefore
argue that there are no significant variations in the temperature of the
X-ray emitting plasma across the angular extent of \snrg.
\par
For comparison purposes with the results for fitting the spectra with
the APEC model, we also fit the extracted spectra
with a non-equilibrium ionization (NEI) plasma model \citep{BLR01}.
Our fitted results with the NEI model -- when compared on the basis of 
the Cash statistics -- do not differ in a significant way from our
fitted results with the APEC model. Furthermore, our fits using the
NEI model include a fitted value for the ionization timescale of 
$\tau\sim10^{13}$ cm$^{-3}$ s for the rims as well as the whole SNR.
The lower limit on our fitted value for the ionization
timescale (at the 90\% confidence level) was 10$^{12}$ cm$^{-3}$ s,
which is consistent with the interpretation that the X-ray emitting
plasma of \snrg\ is in CIE \citep{SH10}. This result is also consistent with our
previous result where we obtained a statistically acceptable fit 
with the APEC model, which assumes CIE conditions for the plasma.
\par
The fitted values of the normalizations using the APEC model for the
different regions of \snrg\ may be used to estimate the values of the 
electron number densities of the X-ray emitting gas at each rim. For the 
APEC model, the normalization
is defined as \begin{equation}
\mbox{Normalization (cm$^{-5}$)} = \frac{10^{-14}}{4 \pi d^2} \int f n_{\rm e} n_{\rm p} dV, 
\end{equation}
where $d$ is the distance to the SNR in cm, $f$ is the volume filling factor, $n_{\rm e}$ and $n_{\rm p}$ are the number densities of the electrons and protons of the region of spectral
extraction in units of cm$^{-3}$ , and $V=\int dV$ is the volume of the region of spectral extraction in units of cm$^3$. Adopting a distance
to \snrg\, of 5.2 kpc based on arguments presented in Sect. 
\ref{PreviousSNRStudiesSection}, $d$ is therefore 1.60$\times$10$^{22}$ cm. Furthermore, if we assume that the volumes of the
regions of spectral extraction are ellipsoids with dimensions of semimajor axis $\times$ semiminor axis $\times$ semiminor axis, then the volumes of the whole SNR, the east rim, the north rim, and the west rim are 1.90$\times$10$^{59}$cm$^3$, 
1.23$\times$10$^{58}$ cm$^3$, 1.71$\times$10$^{58}$ cm$^3,$ and 2.05$\times$10$^{58}$ cm$^3$, respectively. 
Finally, assuming that $n_\mathrm{e}$ = 1.2 $n_\mathrm{p}$ and using our fitted values for the normalizations and the calculated volumes for each region, we estimate the electron densities associated with the whole SNR, the
east rim, the north rim, and the west rim to be 0.29$f$$^{-1/2}$ 
cm$^{-3}$, 0.43$f$$^{-1/2}$ cm$^{-3}$, 0.21$f$$^{-1/2}$ 
cm$^{-3}$, and 0.21$f$$^{-1/2}$ cm$^{-3}$, respectively. We physically interpret these results
to indicate that \snrg\ is expanding into an ambient interstellar medium with approximately constant 
density across its entire azimuth, though the electron density appears to be slightly elevated along
the bright eastern rim of the SNR. Finally, based on the derived electron density for the whole SNR,
we estimate the swept-up mass of the X-ray emitting plasma associated with \snrg\ to be $\sim46f^{-1/2}$ $M_{\rm{\odot}}$.
Additional analysis is required to explore the X-ray properties of this SNR in more detail. 

\section{Summary}

In this paper, we have discussed the properties of the poorly studied Galactic SNR \snrg. Here, we summarize our main findings.
\begin{enumerate}
 \item We have presented the low-frequency observations and flux density measurements of the Galactic SNR \snrg\ using the MWA.
 \item Combining these new MWA observations with the surveys at higher radio continuum frequencies, we discussed the SED of this particular remnant. 
 Synchrotron radio spectral index is estimated to be 0.42$\pm$0.03 and the integrated flux density at 1~GHz to be around 2.7~Jy. 
 \item We also propose that the central point radio source, located apparently inside the SNR shell, is possibly a compact remnant of the supernova explosion with typical steep pulsar SED of 0.99$\pm$0.05. 
 \item We have analyzed the X-ray properties of 
 \snrg\ using an archival pointed observation made of this SNR with {\it XMM-Newton}. We believe that our work describes the first detection and analysis of X-ray emission from \snrg. 
 The spatial distribution of the X-ray emission of \snrg\ as detected by {\it XMM-Newton} broadly matches the spatial distribution of the radio emission: the 
 X-ray morphology of the SNR is shell-like rather than center-filled. The radio-bright eastern and western rims are also readily
 detected in the X-ray while the radio-weak northern and southern rims are weak or absent in the X-ray. We have extracted and fit MOS1+MOS2+PN spectra from the whole SNR as well as the northern, eastern, 
 and western rims of the SNR: by the simultaneous fitting of source and background spectra along with accounting for the
 astrophysical X-ray background, we have successfully fit these extracted spectra with a single thermal model corresponding to an optically thin thermal plasma in collisional ionization equilibrium. 
 The fitted column density toward the SNR is $N_\mathrm{H}\sim0.80\times10^{22}$ $cm^{-2}$ and the fitted temperatures for the spectra for the
 whole SNR as well as the bright rims are all 
 comparable ($kT\sim0.14 - 0.23$ keV). The derived
 electron number densities for the whole SNR and
 the rims are also roughly comparable 
 ($\sim$$0.20f^{-1/2}$ cm$^{-3}$), though the
 electron number density for the eastern rim is slightly larger ($\sim$0.43$f$$^{-1/2}$ cm$^{-3}$). We also
 estimate the swept-up mass of the X-ray emitting
 plasma associated with \snrg\ to be 
 $\sim46f^{-1/2}$ $M$$_{\odot}$.
 \item New high-frequency radio continuum observations (i.e., using the Australia Telescope Compact Array (ATCA), the Australian Square Kilometre Array Pathfinder (ASKAP)) 
 as well as the deeper X-ray observations (e.g., using \textit{XMM-Newton} telescope) will presumably shed more light on the physical properties of this SNR.
\end{enumerate}

\begin{acknowledgements}
We thank the anonymous referee for useful comments that greatly improved the quality of this paper.
This scientific work makes use of the Murchison Radio-astronomy Observatory, operated by CSIRO. We acknowledge the Wajarri Yamatji people as the traditional owners of the Observatory site. 
Support for the operation of the MWA is provided by the Australian Government (NCRIS), under a contract to Curtin University administered by Astronomy Australia Limited. 
We acknowledge the Pawsey Supercomputing Centre, which is supported by the Western Australian and Australian Governments. 
This work is part of the project 176005 ``Emission nebulae: structure and evolution'' supported by the Ministry of Education, Science, and Technological Development of the Republic of Serbia. T.~G.~P. would like to thank Dirk Grupe (Morehead State University), Aviv Brokman (University of Kentucky) and Eric Roemmele (University of Kentucky) for useful discussions about 
Cash statistics.
\end{acknowledgements}


\end{document}